\documentclass[12pt]{article}

\usepackage{epsfig}

\def\balpha{\mbox{\boldmath $\alpha$}}
\input{amssymb.sty}

\begin{document}

\begin{titlepage}

\baselineskip 24pt

\begin{center}

{\Large {\bf Possible Anomalies in Higgs Decay: Charm Suppression 
and Flavour-Violation}}

\vspace{.5cm}

\baselineskip 14pt

{\large Jos\'e BORDES}\\
jose.m.bordes\,@\,uv.es\\
{\it Departament Fisica Teorica, Universitat de Valencia,\\
  calle Dr. Moliner 50, E-46100 Burjassot (Valencia), Spain}\\
\vspace*{.4cm}
{\large CHAN Hong-Mo}\\
h.m.chan\,@\,rl.ac.uk \\
{\it Rutherford Appleton Laboratory,\\
  Chilton, Didcot, Oxon, OX11 0QX, United Kingdom}\\
\vspace{.4cm}
{\large TSOU Sheung Tsun}\\
tsou\,@\,maths.ox.ac.uk\\
{\it Mathematical Institute, University of Oxford,\\
  24-29 St. Giles', Oxford, OX1 3LB, United Kingdom}

\end{center}

\vspace{.3cm}

\begin{abstract}

It is suggested that the Higgs boson may have a branching ratio into the
$c \bar{c}$ mode suppressed by several orders of magnitude compared 
with conventional predictions and in addition some small but
detectable flavour-violating modes such as $b \bar{s}$ and $\tau 
\bar{\mu}$.  The suggestion is based on a scheme proposed and tested
earlier for explaining
the mixing pattern and mass hierarchy of fermions in terms of a 
rotating mass matrix.  If confirmed, the effects would cast new light
on the geometric origin of fermion generations and of the Higgs field
itself.

\end{abstract}

\end{titlepage}

\clearpage

\baselineskip 14pt

Particle physics is, and has been for some time, in the unusual
position of having in the Standard Model an excellent description
of experiment, with as yet no established deviation to give a hint 
of the origin of those very intricate but ad hoc structures built 
into the model.  Fortunately, the situation may soon change with 
the commissioning of the LHC (Large Hadron Collider).  One of the 
first discoveries by the LHC is likely to be the Higgs boson, if it 
exists, and a study of its properties, looking for departures from 
Standard Model predictions, may well provide us with the first 
indications for new physics lying behind and/or beyond. 

It is thus with considerable interest for us to note that a scheme
we have previously suggested to ``explain'' fermion mixing and mass
hierarchy, two outstanding features built into the Standard Model,
may in fact lead to deviations from the standard predictions for
Higgs decay.  Two in particular, namely, the suppression of the
$c \bar{c}$ mode, and the occurence of the flavour-violating modes 
$b \bar{s}$ and $\tau \bar{\mu}$, may be detectable soon after the 
Higgs boson's discovery.  To display how these conclusions result, 
based on a fermion mass matrix rotating with scale, we shall need 
first to outline briefly how this rotation scheme works.  

As already noted, two outstanding features inserted as inputs to 
the Standard Model are the mixing patterns and hierarchical masses 
of quarks and leptons, which remain among the subject's greatest 
mysteries, having as yet no generally accepted explanation and 
accounting for some two-thirds of the model's twenty odd empirical 
parameters.  Among those attempts on offer for explaining the
origin of these features is that of ours which suggests that both 
these two phenomena arise from the change in orientation (rotation) 
with change in scale of the fermion mass matrix in generation space. 
Just as the strength of couplings and mass values can change with 
scale under renormalization, so can, no doubt, the orientation of 
fermion mass matrices.  Indeed, even in the Standard Model as 
conventionally formulated, rotation of fermion mass matrices will 
result as a consequence of mixing from the renormalisation group 
equations \cite{Ramond}.  But, whereas in the Standard Model it is
the pre-assumed fermion mixing which is driving the rotation, the
suggestion now is that it is the rotation instead which is giving 
rise to the mixing of fermions, and as a by-product, also the mass 
hierarchy.

That a rotating fermion mass matrix can give rise to both mixing 
and mass hierarchy is in itself a very simple idea which can easily 
be seen as follows.  One starts with a fermion mass matrix of the 
usual form:
\begin{equation}
m \frac{1}{2}(1 + \gamma_5) + m^{\dagger} \frac{1}{2}(1 - \gamma_5),
\label{m}
\end{equation}
which, following Weinberg \cite{Weinberg}, one can always rewrite
by a relabelling of the singlet right-handed fields, with no change
in physics, in a Hermitian form independent of $\gamma_5$, a form
we shall henceforth adopt.  Suppose now this matrix is factorisable,
meaning that it is of the form:
\begin{equation}
m = m_T \balpha \balpha^{\dagger},
\label{mfact}
\end {equation}
where $\balpha$ is a global (i.e. $x$-independent) vector in
generation space.  We can even suppose that $\balpha$ is
universal and that only the numerical coefficient $m_T$ depends on 
the fermion type (species), i.e.\ whether up-type quarks $(T = U)$, 
down-type quarks $(T = D)$, charged leptons $(T = L)$, or neutrinos 
$(T = N)$.  Obviously, such a mass matrix has only one massive state 
represented by the vector $\balpha$, and zero mixing, i.e. only 
the identity matrix as the mixing matrix.  This is not unattractive 
as a starting point at least for quarks, as has occurred already a 
long time ago to many people \cite{Fritsch},
but is clearly insufficiently realistic in detail.

However, if one now says that the vector $\balpha$ rotates with
changing scale $\mu$, as proposed, then the situation becomes very
interesting.  All quantities now depend on the scale $\mu$ and one 
has to specify at which scale the mass or state vector of each
particle is to be measured.  Suppose one follows the usual convention
and define the mass of each particle as that measured at the scale 
equal to its mass, we find then that mixing and mass hierarchy would
immediately result.  This is apparent already in a situation where one 
takes account only of the two heaviest states in each species, e.g. 
$t, c$ in up-type quarks $(U)$, $b, s$ in down-type quarks $(D)$, which 
``planar approximation'', as we shall see, is already good enough for 
most of the considerations in this paper.

\begin{figure} [ht]
\centering
\input{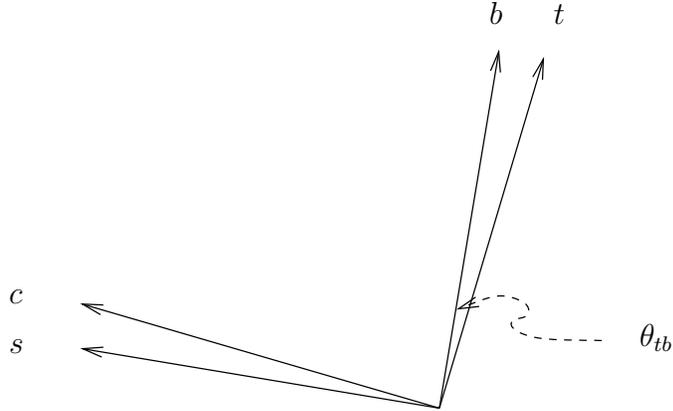}
\caption{Mixing between up and down fermions from a rotating mass matrix.}  
\label{UDmix}
\end{figure}

By (\ref{mfact}) then, taking for the moment $\balpha$ to be 
real and $m_T$ constant for simplicity, we would have $m_t = m_U$ 
as the mass of $t$ and the eigenvector $\balpha (\mu = m_t)$ as 
its state vector ${\bf v}_t$.  Similarly, we have $m_b = m_D$ as the 
mass and $\balpha (\mu = m_b)$ as the state vector ${\bf v}_b$ of 
$b$.  Next, the state vector ${\bf v}_c$ of $c$ must be orthogonal to 
${\bf v}_t$, $c$ being by definition an independent quantum state to 
$t$.  Similarly, the state vector ${\bf v}_s$ of $s$ is orthogonal to 
${\bf v}_b$.  So we have the situation as illustrated in Figure 
\ref{UDmix}, where the vectors ${\bf v}_t$ and ${\bf v}_b$ are not 
aligned, being the vector $\balpha (\mu)$ taken at different 
values of its argument $\mu$, and $\balpha$ by assumption rotates.  
This gives then the following CKM mixing (sub)matrix in the ``planar 
approximation'' for the two heaviest states:
\begin{equation}
\left( \begin{array}{cc} V_{cs} & V_{cb} \\ V_{ts} & V_{tb} \end{array}
   \right) = \left( \begin{array}{cc} \langle {\bf v}_c|{\bf v}_s \rangle 
                             & \langle {\bf v}_c|{\bf v}_b \rangle \\
                               \langle {\bf v}_t|{\bf v}_s \rangle 
                             & \langle {\bf v}_t|{\bf v}_b \rangle
             \end{array} \right )
           = \left( \begin{array}{cc} \cos \theta_{tb} & \sin \theta_{tb} \\  
                -\sin \theta_{tb} & \cos \theta_{tb} \end{array} \right),    
\label{UDmixing}
\end{equation}
which is no longer the identity; hence mixing.  Clearly, the same 
argument can be extended to the three generation case and, just by 
virtue of rotation, a full, non-trivial CKM matrix would result.
The same again when applied to leptons will lead to a nontrivial MNS 
mixing matrix and give rise to neutrino oscillations.    

\begin{figure} [ht]
\centering
\includegraphics{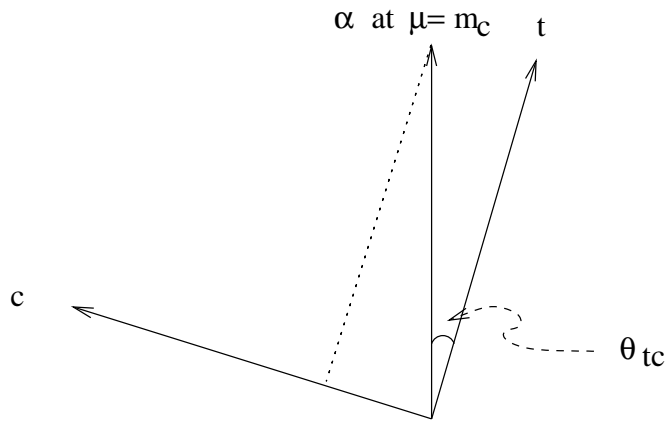}
\caption{Masses for lower generation fermions from a rotating mass matrix
via the ``leakage'' mechanism.}  
\label{alphaU}
\end{figure}

Next, what about hierarchical masses?  From (\ref{mfact}), it follows 
that ${\bf v}_c$ must have zero eigenvalue at $\mu = m_t$.  But this 
value is not to be taken as the mass of $c$ which we agreed has to be 
measured at $\mu = m_c$.  In other words, $m_c$ is instead to be taken 
as the solution to the equation:
\begin{equation}
\mu = \langle {\bf v}_c|m(\mu)|{\bf v}_c \rangle 
    = m_U |\langle {\bf v}_c|\balpha (\mu) \rangle|^2.
\label{solvmc}
\end{equation}
A nonzero solution exists since $\balpha$ by assumption rotates
so that at $\mu \neq m_t$, it would have rotated to some direction
different from ${\bf v}_t$, as illustrated in Figure \ref{alphaU}, 
and acquired a component, say $\sin \theta _{tc}$, in the direction 
of ${\bf v}_c$ giving thus:
\begin{equation}
m_c = m_t \sin^2 \theta_{tc},
\label{m_c}
\end{equation}
which will be small if the rotation is not too fast.  Clearly, similar
arguments when applied to the 3 generation case would lead to a small 
nonzero mass for $u$.  They could be repeated also for the down-type 
quarks and charged leptons, and with a further twist supplied by the
see-saw mechanism, for neutrinos as well.  Hence we have the mass 
hierarchy as claimed, with the two lowest generations of each type 
(or species) acquiring their small nonzero masses as if by ``leakage'' 
from the heaviest generation, again simply by virtue of the rotation.  

\begin{figure}
\centering
\hspace*{-3.5cm}
\input{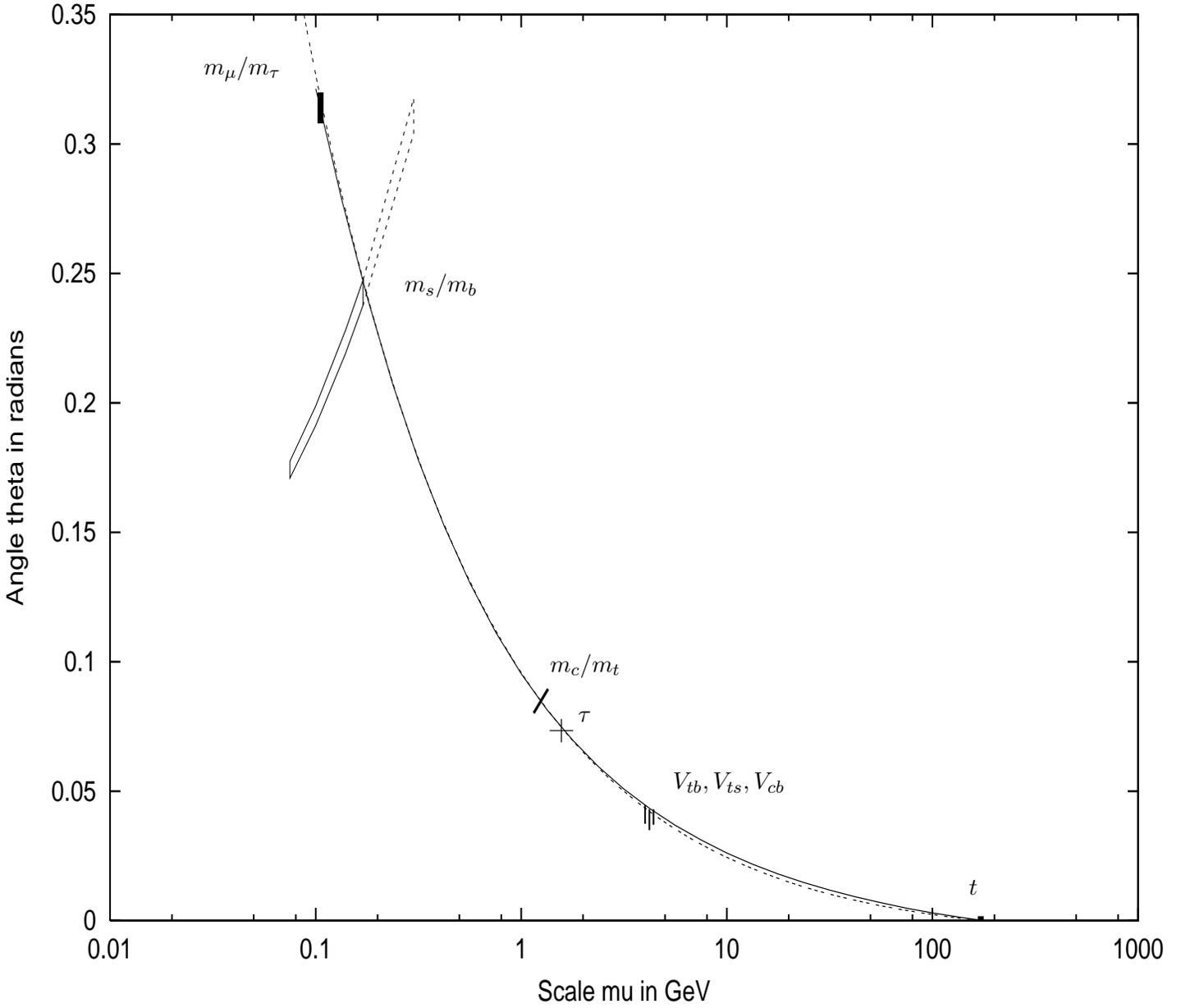}
\caption{The rotation angle changing with scale as extracted from data on
mass ratios and mixing angles and compared with the best fit to the data
(dashed curve) and the earlier calculation by DSM 
(full curve) \cite{phenodsm}, in the planar approximation.}
\label{planarplot}
\end{figure}

The examples in the preceding paragraphs show that both mixing and
mass hierarchy will naturally follow as consequences of rotation,
(For a more detailed discussion, see e.g. \cite{strongcp}) but
will the results be anything like those experimentally observed?  
This question was examined in an earlier paper \cite{cevidsm} where
the rotation angles needed for explaining the mixing parmeters and
mass ratios obtained in experiment were plotted against the scale
$\mu$.  If the explanation is correct, then the angles so extracted
should all lie on one smooth curve.  The figure so obtained in the
``planar approximation'' is reproduced in Figure \ref{planarplot}
here as it will be of use to us later.  One sees that the idea is 
well-sustained.  A full analysis taking account of all three 
generations has also been performed which, though still entirely 
consistent with the rotation hypothesis, does not add too much to 
Figure \ref{planarplot}, given the imprecision of neutrino data yet 
available \cite{cevidsm}.  We have not updated the
fit in Figure \ref{planarplot} with more recent data since
 there have been only small changes 
for most of the
quantities needed, which will not materially affect the fit and any
conclusions that will be drawn from it later.  
\footnote{The only exception is the $s$ mass, which is, however, 
always a little ambiguous in the way it is extracted from data and 
makes it quite unclear to what it is to be compared in our rotation 
picture.  And since its cited value has varied considerably over the 
years and carries always a large error, we thought it best just to 
ignore it in our fit.}   

Suppose we take seriously the proposal that rotation does offer a 
feasible explanation for fermon mixing and hierarchy.  Then one has
to pose oneself the following two questions.  First, can one device
a physically viable and theoretically consistent model in which the 
rotation required by experiment is reproduced?  Secondly, apart from 
explaining existing data, are there any new predictions which can be 
tested by experiment?  The answer to the first has been and is still
being diligently pursued.  So far two quite detailed models have been 
constructed, one phenomenological \cite{phenodsm,genmixdsm} which 
fits experiment quite well but is theoretically incomplete, while the 
other starts from a much sounder theoretical base but has not yet been
sufficiently studied to show if it yields a good fit to experiment 
\cite{prepsm}.  However, the details of these models need not bother 
us in this paper which is devoted to the second question and, as we 
shall see, the predicted anomalies in Higgs decay we are discussing 
all depend only on the general concept of mass matrix rotation, not 
on any details of either of the two models proposed. 

To investigate Higgs decay, we shall need its Yukawa coupling.  One
obvious possibility which will give the required factorisable mass 
matrix (\ref{mfact}) is the following:
\begin{equation}
{\cal A}_{YK} = \rho_T \balpha \bar{\psi}_L \phi \psi_R 
{\balpha}^{\dagger} + {\rm h.c.}
\label{Yukawa}
\end{equation}
and indeed, this is the Yukawa coupling obtained in our theoretical 
model \cite{prepsm}.  Choosing the gauge in which $\phi$ points 
in the up direction and is
real, then expanding the remaining real component about its minimum
value $\zeta_W$, thus:
$\zeta_W + H$ , we obtain to zeroth order the 
fermion mass matrix as in (\ref{mfact}) with $m_T = \rho_T \zeta_W$, 
and to first order the coupling matrix of the Higgs boson to the 
fermions as:
\begin{equation}
Y = \rho_T \balpha \balpha^{\dagger}.
\label{Hcoup}
\end{equation}

Superficially, this result looks familiar, namely that the fermion
mass and Higgs coupling are proportional to each other.  However, as 
usually meant, the proportionality is between the mass and the Higgs 
coupling of each of the fermions individually, namely that $m_i = 
\zeta_W y_i$, with $i$ denoting the individual fermion state, but 
here it is a proportionality between matrices: $m = \zeta_W Y$.  
And both these matrices rotate.  Having seen above that the rotation 
of the mass matrix $m$ alone already leads to intriguing consequences, 
we shall not be surprised that here too the rotation of the Higgs 
coupling matrix $Y$ will give some new and interesting results. 

For the present, let us concentrate just on Higgs boson decay into
fermion-antifermion pairs, where the relevant couplings are to be 
extracted from (\ref{Hcoup}).  Since (\ref{Hcoup}) depends on the
vector $\balpha$ which in turn depends on the scale $\mu$, 
i.e. rotates, we have first to specify at what scale to take this
$\balpha$ and what value it will then possess.  For Higgs
decay, the conventional wisdom is to take the scale at the Higgs 
mass $M_H$, a convention we adopt.  Now, $\balpha$ rotates 
with scale according, presumably, to some renormalization group 
equation, which is a differential equation in $\mu$ telling us how 
$\balpha$ varies with changes in $\mu$.  Therefore, to evaluate 
$\balpha$ at $\mu = M_H$ as prescribed, we shall need first to 
calibrate $\balpha$ by an initial condition.  We propose to do so 
as follows.  At the energy scale $\mu = 2 m_f$, i.e. at the threshold 
of production of the $f \bar{f}$-pair, we say the fermions are at 
rest so that the Higgs coupling there should be proportional to the 
fermion mass, thus: 
\begin{equation}
\langle {\bf v}_f|Y(\mu = 2 m_f)|{\bf v}_f \rangle 
= \rho_T |\langle {\bf v}_f|\balpha(\mu = 2 m_f)\rangle|^2   
= m_f/\zeta_W.
\label{acalibg}
\end{equation}
Recall now that previously, when extracting the fermion masses from 
the rotating mass matrix, we had put:
\begin{equation}
m_f = \zeta_W \rho_T |\langle {\bf v}_f|\balpha(\mu = m_f)|^2,
\label{acalibm}
\end{equation}
though without stating so explicitly.  We see that we have now in
(\ref{acalibg}) calibrated our $\balpha$ differently; namely 
we have shifted the scale by an overall factor $2$ compared with 
(\ref{acalibm}), but otherwise there is no difference between the 
two calibrations.  This means first that our calibration here is
self-consistent, being independent of the fermion $f$ with which we 
choose to calibrate.  Secondly, it also means that the value we seek
for the factor $\balpha_H \balpha_H^{\dagger}$ for Higgs
decay at $\mu = M_H$, which we may interpret as the ``state tensor''
of the Higgs state, can just be read off, for phenomenological
purposes, from the trajectory for $\balpha(\mu)$ obtained in 
\cite{cevidsm} by fitting fermion masses and mixing parameters, i.e. 
apart from a change in scale by a factor 2, or just a shift of the 
origin by $\ln 2$ there on the $\ln \mu$ plot.  That there should 
be such a shift in scale between the calibrations (\ref{acalibg}) 
and (\ref{acalibm}) is not surprising, given that we are dealing 
here with two different types of processes, one type, Higgs decays, 
involving two fermions in the final states, and the other type, 
fermion propagation as concerns masses and mixing angles, involving 
only a single fermion.  Nor is this shift, as we shall see later, 
of much qualitative significance. 

The conclusions of the preceding paragraph then give the coupling 
for Higgs decaying into an $f \bar{f}$ pair as:
\begin{equation}
A(H \rightarrow f \bar{f}) = \rho_T |{\bf v}_f.\balpha_H|^2.
\label{ampHdecay}
\end{equation}

Specialising now to the $c \bar{c}$ and $b \bar{b}$ modes, we have:
\begin{equation}
\frac{\Gamma(H \rightarrow c \bar{c})}{\Gamma(H \rightarrow b \bar{b})}
   \sim \frac{\rho_U^2}{\rho_D^2}\frac{|{\bf v}_c.\balpha_H|^4}
      {|{\bf v}_b.\balpha_H|^4},
\label{BRcoverb}
\end{equation}
where $\rho_U/\rho_D \sim m_t/m_b \sim 40.8$, in which estimate, as
in all other estimates of similar nature in this paper, we have taken
the values of measured quantities from \cite{databook}.  Since only 
the two heaviest generations are involved, the planar approximation
of \cite{cevidsm} should be enough for the present exploration.  In
that case, the state vectors ${\bf v}_c, {\bf v}_b$ as well as the
vector $\balpha_H$ are each given just by an angle $\theta$,
the value of which can be read off at the appropriate values of
$\mu$ from Figure \ref{planarplot}, or else calculated with the 
simple formula obtained there as the best fit.  Remembering now the 
shift in scale, we note that the value of $\theta_H$ corresponding 
to $\balpha_H$ is to be the value of $\theta$ at $\mu = M_H/2$ 
on this plot.  The value of $M_H$ is bounded by present experiment 
to be above 115 GeV and is generally assumed to be below the $t$ 
mass.  Taking then as example $M_H = 150$ GeV, we obtain for $\mu 
= 75$ GeV the value $\theta_H \sim 0.0040$, which gives then:
\begin{equation}
|{\bf v}_c.\balpha_H| = \sin (\theta_H - \theta_t) \sim 0.0040;
\ \ \ 
|{\bf v}_b.\balpha_H| = \cos (\theta_b - \theta_H) \sim 0.9993,
\label{estcHbH}
\end{equation}
and the branching ratio (\ref{BRcoverb}) as:
\begin{equation}
\frac{\Gamma(H \rightarrow c \bar{c})}{\Gamma(H \rightarrow b \bar{b})}
   \sim 4.3 \times 10^{-7}.
\label{preBRcoverb}
\end{equation}
This is nearly five orders of magnitude less than the conventional 
prediction of this branching ratio $\sim m_c^2/m_b^2 \sim 0.09$.  We 
note in passing that had we not shifted the scale by a factor $2$ as 
we did above, we would have obtained $\theta_H \sim 0.00059$ instead, 
which would have suppressed the $c \bar{c}$ mode even further.

The reason for this suppression for $c \bar{c}$ mode in the rotation
scenario is clear.  The factor $m_c^2$ which occurs in the standard
conventional formula for this branching ratio is given in the
rotation picture as $m_t^2 \sin (\theta_c - \theta_t)^4$, where
$\sin (\theta_c - \theta_t)$ represents the component of $\balpha$
taken at $\mu = m_c$ in the direction of ${\bf v}_c$, i.e. orthogonal
to ${\bf v}_t$.  Now, in the present scenario, we are instead to take
the value of $\balpha$ at $\mu = M_H$ which has a value close to 
${\bf v}_t$, since $M_H$ is close to $m_t$, and has therefore a very 
small component in the direction of the state vector ${\bf v}_c$ of 
$c$.  Hence we have the suppression.  It comes about as a consequence, 
first, of rotation and, secondly of our insistence in evaluating the 
Higgs coupling (or state tensor) at the scale $\mu = M_H$ thought 
appropriate for Higgs decay in accordance to conventional wisdom.

Clearly, the same arguments will apply to other $f \bar{f}$ modes for
$f$ belonging to the two lighter generations, giving, for example,
for Higgs decaying into $s \bar{s}$ and $\mu \bar{\mu}$ the branching 
ratio over the main mode $b \bar{b}$ estimates of order respectively 
$1.8 \times 10^{-6}$ and $2.9 \times 10^{-6}$, again for $M_H = 150\ 
{\rm GeV}$.  But the conventional predictions for these modes being 
in themselves rather small, namely $\sim 6 \times 10^{-4}$ and 
$\sim 6.4 \times 10^{-4}$ respectively, a further suppression of 
these may not be so readily seen.  However, a suppression of the 
$c \bar{c}$ mode from the respectable standard estimate of $\sim 0.09$ 
may be noticeable soon after the Higgs boson is found. 

Perhaps more surprisingly, similiar considerations to those above 
in the same rotation scenario will lead also to flavour-violations in 
Higgs decay.  The Higgs coupling (or the Higgs boson ``state tensor'') 
$\balpha_H \balpha_H^{\dagger}$ is in general nondiagonal 
when sandwiched between the fermion state vectors ${\bf v}_f$ and 
${\bf v}_{f'}$.  It gives thus nonzero rates to flavour-violating 
modes, but the branching fractions for these are generally quite small.  
Take, for example, $H \rightarrow \tau \bar{\mu}$.  Since only the two 
heaviest generation leptons are involved, we may again work with the 
planar approximation of Figure \ref{planarplot} from which we read the 
value of $\theta_{\tau} \sim 0.0698$.  Hence, and from the value of 
$\theta_H \sim 0.0040$ given previously, we have:
\begin{equation}
\frac{\Gamma(H \rightarrow \tau \bar{\mu})}
   {\Gamma(H \rightarrow b \bar{b})} \sim \frac{m_{\tau}^2}{m_b^2}
   \frac{\cos^2 \theta_{H \tau} \sin^2 \theta_{H \tau}}
   {\cos^4 \theta_{H b}} \sim 7.7 \times 10^{-4}.
\label{BRtaumu}
\end{equation}
which, being so distinctive though small, may eventually be seen.
Another possibility to try may be $H \rightarrow b \bar{s}$ with
a slightly larger branching ratio over the dominant $b \bar{b}$
mode of about $1.48 \times 10^{-3}$ but perhaps is harder to identify.

This prediction of flavour-violation in Higgs decay is at first sight 
quite disturbing, as the effect may propagate and give rise 
to flavour-violation elsewhere where it is already very strongly 
bounded by experiment.  We therefore examine the following two cases 
which look to us most dangerous, namely $m_{B^0_{sH}} - m_{B^0_{sL}}$,
the mass difference between the two neutral $b\bar{s}, s \bar{b}$ 
bound states, and the rate for the decay mode $\tau^{\pm} \rightarrow 
\mu^+ \mu^- \mu^{\pm}$, in both of which the rotation scenario is 
likely to give the largest effects and the existing experimental 
bounds are already quite strict.  

For the first quantity, following standard procedure, given e.g. 
in \cite{Donohue}, we obtain the estimate for the contribution 
from Higgs exchange:
\begin{equation}
m_{B^0_{sH}} - m_{B^0_{sL}} \sim \frac{\rho_D^2}{M_H^2}
   |\balpha_H.{\bf v}_b|^2 |\balpha_H.{\bf v}_s|^2 F_{B_s}
m_{B_s},
\label{DeltamBs}
\end{equation}
where $\rho_D \sim m_b/\phi_0 \sim 0.0171$, $\phi_0$ being the vacuum
expectation value of the Higgs field $\sim 246\ {\rm GeV}$, $m_{B_s}
\sim 5366\ {\rm MeV}$, and
$F_{B_s}$ is a factor related to the decay constant $f_{B_s}$ of the
$B_s$ meson, which, if assumed to be similar to the analogous factor
$F_K = f^2_K/3$ occurring in the formula for the FCNC estimate for
the $K_L - K_S$ mass difference, gives $F_{B_s}
\sim 5 \times 10^{-3}\ {\rm GeV^2}$ \footnote{The factor $F_{B_s}$
here differs from the usual factor in the K mass difference from FCNC
in that the scalar current is involved here, but using a result of
\cite{Bebanson} applied to this case, the resulting value
is not very different.}
This gives an estimate for the above new contribution 
as:
\begin{equation}
m_{B^0_{sH}} - m_{B^0_{sL}} \sim 4.8 \times 10^{-10}\ {\rm MeV}
\label{DeltamBse}
\end{equation}
which is more than an order of magnitude below the experimentally
measured value of $117 \times 10^{-10}\ {\rm MeV}$.  Given the
difficulty of making accurate theoretical predictions for such
hadronic quantities, the Higgs contribution is unlikely to be
noticeable at present and thus causes no problem, but it may be 
something to look for in future when experimental measurement and 
theoretical interpretation both continue to improve.

For $\tau^{\pm} \rightarrow \mu^+ \mu^- \mu^{\pm}$ decay, one can
make use of its similarity to $\tau \rightarrow \nu_{\tau} \mu
\nu_{\mu}$ decay when one neglects in both cases the masses of the
final particles to give the ratio of their total rates as:
\begin{equation}
\frac{\Gamma(\tau^{\pm} \rightarrow \mu^+ \mu^- \mu^{\pm})}
   {\Gamma(\tau \rightarrow \nu_{\tau} \mu \nu_{\mu})}
\sim \frac{1}{64} \left( \frac{\rho_L}{g_W} \right)^4
   \left( \frac{M_W}{M_H} \right)^4 |\balpha_H.{\bf v}_\tau|^2
   |\balpha_H.{\bf v}_\mu|^6,
\label{tauto3mu}
\end{equation}
where $\rho_L \sim m_\tau/\phi_0 \sim 7.2 \times 10^{-3}$, $g_W 
\sim 0.23$ is the coupling and $m_W \sim 80.4\ {\rm GeV}$ the mass 
of the $W$ boson.  Feeding in then the values for $\balpha_H.
{\bf v}_\tau$ and $\balpha_H.{\bf v}_\mu$ obtained earlier 
when deriving (\ref{BRtaumu}), we obtain an estimate for the ratio 
(\ref{tauto3mu}) of about $1 \times 10^{-16}$, or a branching ratio 
over total of about $2 \times 10^{-17}$, which is way below the 
present experimental bound of around $3 \times 10^{-8}$.  This very
small value of the estimated rate for (\ref{tauto3mu}) comes partly 
from the smallness of the coupling strength $\rho_L$ itself but 
partly also from the smallness of its branching fraction into $\mu$, 
namely $|\balpha_H.{\bf v}_\mu|$. 

A similar analysis applied to other mass differences and to other
flavour-violating decays give even smaller estimates.  It appears 
thus likely
that the prediction of flavour-violating modes in Higgs decay is 
not in any contradiction with existing bounds on 
flavour-violation 
elsewhere.  We hope to give a more detailed and systematic account 
of the analysis and its result in a separate report.

In charm suppression and flavour-violating modes in Higgs decay,
we seem thus to have identified two possible deviations from the
Standard Model which can be tested soon after the Higgs boson's
experimental discovery.  We say ``possible'' only, because we do 
not have, as yet, a fully developed rotational model which allows 
us to derive the required results from first principles.  As it 
is done here, based merely on the general concept of rotation, 
we arrived at the above results by extending the logic previously 
tested in deriving the mixing pattern and mass hierarchy 
just one step 
further to decay processes.  It is thus not clear at this stage 
that the same logic can be consistently extended to cover all 
possible processes involving an arbitrary number of particles, as 
would be necessary if the logic is correct, although of course we 
have explored some other examples and found yet no inconsistency.
Nevertheless, the experimental discovery of the Higgs boson being, 
we hope, imminent, we believe this possibility worth airing, given 
that, if proved correct, it will open a valuable window into the 
foundations of the Standard Model.  This is because the rotation 
idea has the potential of explaining how fermion mixing and mass 
hierarchy arise, which are two of the most mysterious assumptions, 
and most costly in terms of parameters, built into the Standard 
Model.  Besides, the prediction, if confirmed, may also teach us
a great deal about the possible origin both of fermion generations 
and of the Higgs scalar fields, two fundamental features which, 
in the rotation models \cite{prepsm} are both intimately connected 
with fermion mixing and fermion mass hierarchy, and are both given 
a geometrical significance, with fermion generations arising as 
dual colours and Higgs scalar fields arising as frame vectors in 
internal symmetry space.

\end{document}